\documentclass[12pt]{iopart}

\usepackage{iopams}
\usepackage{graphicx}
\usepackage[breaklinks=true,colorlinks=true,linkcolor=blue,urlcolor=blue,citecolor=blue]{hyperref}

\begin{document}

\title{Relaxation of Rabi Dynamics in a Superconducting Multiple-Qubit Circuit}

\author{Deshui Yu$^{1}$, Leong Chuan Kwek$^{1,2,3,4}$, \& Rainer Dumke$^{1,5}$}

\address{$^{1}$Centre for Quantum Technologies, National University of Singapore, 3 Science Drive 2, Singapore 117543, Singapore}

\address{$^{2}$Institute of Advanced Studies, Nanyang Technological University, 60 Nanyang View, Singapore 639673, Singapore}

\address{$^{3}$National Institute of Education, Nanyang Technological University, 1 Nanyang Walk, Singapore 637616, Singapore}

\address{$^{4}$MajuLab, CNRS-UNS-NUS-NTU International Joint Research Unit, UMI 3654, Singapore}

\address{$^{5}$Division of Physics and Applied Physics, Nanyang Technological University, 21 Nanyang Link, Singapore 637371, Singapore}

\ead{rdumke@ntu.edu.sg}

\begin{abstract}
We investigate a superconducting circuit consisting of multiple capacitively-coupled charge qubits. The collective Rabi oscillation of qubits is numerically studied in detail by imitating environmental fluctuations according to the experimental measurement. For the quantum circuit composed of identical qubits, the energy relaxation of the system strongly depends on the interqubit coupling strength. As the qubit-qubit interaction is increased, the system's relaxation rate is enhanced firstly and then significantly reduced. In contrast, the inevitable inhomogeneity caused by the nonideal fabrication always accelerates the collective energy relaxation of the system and weakens the interqubit correlation. However, such an inhomogeneous quantum circuit is an interesting test bed for studying the effect of the system inhomogeneity in quantum many-body simulation.
\end{abstract}

\section{Introduction}

Owing to the fascinating properties such as flexibility, tunability, scalability, and strong interaction with electromagnetic fields, superconducting Josephson-junction circuits provide an outstanding platform for quantum information processing (QIP)~\cite{Science:Devoret2013,RPP:Wendin2017}, quantum simulation of many-body physics~\cite{NatCommun:Barends2015,PhysRevA.95.042330,SciRep:Lamata2017}, and exploring the fundamentals of quantum electrodynamics in and beyond the ultrastrong-coupling regime~\cite{NatPhys:Niemczyk2010,NatPhys:Yoshihara2017}. Additionally, hybridizing these solid-state devices with the atoms may enable the information transfer between macroscopic and microscopic quantum systems~\cite{PRA:Yu2016-1,SciRep:Yu2016,PRA:Yu2016-2,QST:Yu2017,PRA:Yu2017,ProcSPIE:Hufnagel2017}, where the superconducting circuits play the role of rapid processor while the atoms act as the long-term memory. Nonetheless, the strong coupling to the environmental noise significantly limits the energy-relaxation ($T_{1}$) and dephasing ($T_{2}$) times of superconducting circuits~\cite{PRL:Astafiev2004,PRL:Yoshihara2006,PRA:Koch2007}.

Recently, there has been focus on large-scale QIP~\cite{PRL:Song2017,npj:Lu2017}. Several network schemes have already been demonstrated in experiments: one-dimensional spin chain with the nearest-neighbor interaction~\cite{Nature:Kelly2015,Nature:Barends2016}, two-dimensional lattice with quantum-bus-linked qubits~\cite{NatCommun:Corcoles2015,NatCommun:Riste2015}, multiple artificial atoms interacting with the same resonator~\cite{PRL:Song2017}, and many cavities coupled to a superconducting qubit~\cite{NewJPhys:Yang2016}. However, only little attention has been paid to the quantum circuit composed of nearly identical superconducting qubits, where an arbitrary individual directly interacts with others and, in addition, all qubits are biased by the same voltage, current, or magnetic bias and are exposed to the same fluctuation source. Studying such a multiple-qubit architecture is of importance to superconducting QIP network. For one thing, the nearest- or next-nearest-neighbour-coupling approximation, which is commonly employed in scalable superconducting schemes~\cite{NewJPhys:Wallquist2005,PRB:Storcz2005,PRB:Richer2016}, does not always hold the truth in realistic systems. For another, transferring the quantum information from one processor to another over a long distance, which relies on the long-range interqubit coupling, is essential for cluster quantum computing~\cite{RepMathPhys:Nielsen2006} and quantum algorithms~\cite{Nature:DiCarlo2009}. Moreover, in a large-scale network composed of strong or ultrastrong interacting qubits, the energy relaxation and dephasing of one qubit strongly influence the dynamics of others and this fluctuation can be rapidly boosted and affect more qubits. Such a collective dissipation may significantly degrade the information transfer fidelity between two remotely separated qubits. However, to our best knowledge, this collective dissipation of a large ensemble of directly-coupled superconducting qubits has not been explored before.

Here we investigate the collective relaxation effect in a multiple-qubit circuit, where all charge qubits are biased by the same voltage source, capacitively coupled via linking all islands together, and influenced by the same noise source. The system's Rabi oscillation is numerically studied in detail. It is shown that for the homogeneous system with identical qubits, the interqubit coupling strongly modifies the collective relaxation of quantum circuit. In contrast, the nonideal-fabrication-induced extra inhomogeneity always enhances the system's relaxation rate and also provides a platform for studying many-body localization.

\section{Results}

\subsection{Physical model}

\begin{figure}
\centering
\includegraphics[width=12.0cm]{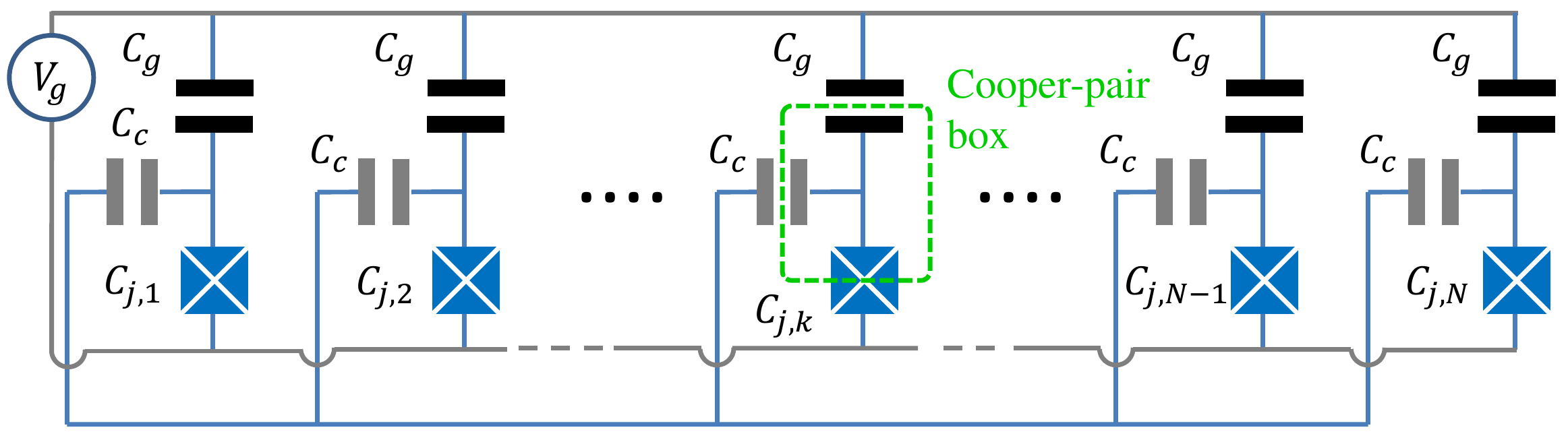}\\
\caption{Multiple-qubit scheme. $N$ charge qubits are biased by the voltage source $V_{g}$ via the identical gate capacitors $C_{g}=300$ aF. In this work, we set $N=10$. All Cooper-pair boxes are linked together via the identical coupling capacitors $C_{c}$. The inhomogeneity of the system only arises from the nonidentical Josephson junctions. $C_{j,k}$ and $E_{j,k}$ with $k=1,\ldots,N$ denote the self-capacitances and Josephson energies of different Josephson junctions. The corresponding mean values are $C_{j}=30$ aF and $E_{J}=2\pi\hbar\times3$ GHz.}\label{Fig1}
\end{figure}

We consider $N$ capacitively-coupled single-Josephson-junction charge qubits as shown in figure~\ref{Fig1}. A voltage source $V_{g}$ biases the array of superconducting islands via the identical gate capacitors $C_{g}$. All Cooper-pair boxes are linked together via the identical coupling capacitors $C_{c}$, where we have applied the fact that the practical inhomogeneities of capacitors $C_{g}$ and capacitors $C_{c}$ are much smaller than that of Josephson junctions.

For the $k$-th ($k=1,\ldots,N$) charge qubit, we define the self-capacitance of Josephson junction, the total capacitance, the charging energy, and the Josephson energy as $C_{j,k}$, $C_{\Sigma,k}=C_{g}+C_{j,k}+C_{c}$, $E_{C,k}=\frac{(2e)^{2}}{2C_{\Sigma,k}}$, and $E_{J,k}$, respectively. The total excess charge $Q_{k}$ of the $k$-th box distributes on capacitor plates of $C_{g}$, $C_{j,k}$, and $C_{c}$ that are involved in the Cooper-pair box. The corresponding charges are defined as $Q_{g,k}$, $Q_{j,k}$, and $Q_{c,k}$, respectively, and we have
\begin{equation}
Q_{k}=Q_{g,k}+Q_{j,k}+Q_{c,k}.
\end{equation}
According to Kirchhoff's circuit laws and charge conservation, one obtains
\begin{eqnarray}
&&\textstyle\frac{Q_{g,1}}{C_{g}}-\frac{Q_{j,1}}{C_{j,1}}=\frac{Q_{g,2}}{C_{g}}-\frac{Q_{j,2}}{C_{j,2}}=\ldots=-V_{g},\\
&&\textstyle\frac{Q_{c,1}}{C_{c}}-\frac{Q_{j,1}}{C_{j,1}}=\frac{Q_{c,2}}{C_{c}}-\frac{Q_{j,2}}{C_{j,2}}=\ldots,\\
&&Q_{c,1}+Q_{c,2}+\ldots+Q_{c,N}=0.
\end{eqnarray}
The total charging energy of the system is then given by
\begin{eqnarray}
\nonumber E_{ch}&=&\textstyle\sum_{k}\left(\frac{Q^{2}_{g,k}}{2C_{g}}+\frac{Q^{2}_{j,k}}{2C_{j,k}}+\frac{Q^{2}_{c,k}}{2C_{c}}+V_{g}Q_{g,k}\right)\\
&=&\textstyle\sum_{k}E_{C,k}\left(N_{k}-N_{g}\right)^{2}+\left[\sum_{k}\frac{E_{C,k}}{\sqrt{NV}}\left(N_{k}-N_{g}\right)\right]^{2},
\end{eqnarray}
where $N_{k}=-\frac{Q_{k}}{(2e)}$ denotes the number of excess Cooper pairs in the $k$-th box and $N_{g}=\frac{C_{g}V_{g}}{(2e)}$ is the gate-charge bias. We have also defined
\begin{equation}
V=\textstyle\frac{(2e)^{2}}{2C_{c}}-\frac{1}{N}\textstyle\sum_{k}E_{C,k},
\end{equation}
which is positive and denotes the difference between the electrostatic energy of a Cooper pair in $C_{c}$ and the mean value of charging energies of different Cooper-pair boxes. Adding the tunneling energies of Cooper pairs into $E_{ch}$~\cite{Book:Dittrich1998}, the system's Hamiltonian is derived as $H=H_{0}+H_{1}$, where
\begin{eqnarray}
H_{0}&=&\textstyle{\sum_{k}\left[\frac{E_{C,k}}{2}(1-2N_{g})\sigma^{(k)}_{z}-\frac{E_{J,k}}{2}\sigma^{(k)}_{x}\right]},\label{H1a}\\
H_{1}&=&\textstyle{\left[\sum_{k}\frac{E_{C,k}}{2\sqrt{NV}}(\sigma^{(k)}_{z}+1-2N_{g})\right]^{2}},\label{H1b}
\end{eqnarray}
in the two-state approximation. The $x$- and $z$-components of the Pauli operator are given by $\sigma^{(k)}_{x}=(|0\rangle\langle1|)_{k}+(|0\rangle\langle1|)_{k}$ and $\sigma^{(k)}_{z}=(|1\rangle\langle1|)_{k}-(|0\rangle\langle0|)_{k}$. $|0\rangle_{k}$ and $|1\rangle_{k}$ represent the absence and presence of a single excess Cooper pair in the $k$-th island.

The multiple-qubit system operates in the charging limit of $E_{C,k}\gg E_{J,k}$. $H_{0}$ gives the total energy of free charge qubits while $H_{1}$ corresponds to the interqubit interaction energy. It is seen that besides a constant, the qubit-qubit coupling leads to the linear terms of $(1-2N_{g})\sigma^{(k)}_{z}$, which may be involved into $H_{0}$, and the quadratic $zz$-interaction terms of $\sigma^{(k_{1})}_{z}\sigma^{(k_{2})}_{z}$ that are diagonal in the charge number basis. In the limit of $C_{c}\sim0$, we have $V\rightarrow\infty$, for which $H_{1}\sim0$ and charge qubits become independent with each other. As $C_{c}$ is increased, $E_{C,k=1,\ldots,N}$ go down. In this work, we restrict $C_{c}$ within the range satisfying the charging limit condition. Due to the assumption of identical gate and coupling capacitors, the system's inhomogeneity completely comes from the nonidentical Josephson junctions.

\subsection{Homogeneous Circuit}

We first consider the homogeneous system, where all Josephson junctions are identical, i.e., $C_{j,k}=C_{j}$, $E_{C,k}=E_{C}$, and $E_{J,k}=E_{J}$. It is easy to obtain $\frac{V}{E_{C}}=\frac{C_{g}+C_{j}}{C_{c}}$ and $H$ can be simplified as
\begin{equation}\label{H2}
\textstyle\frac{H}{N}=\frac{E^{2}_{C}}{4V}\left[\frac{J_{z}}{N}+\Xi(V,N_{g})\right]^{2}-\frac{E_{J}}{2}\frac{J_{x}}{N},
\end{equation}
by defining the collective operators $J_{z}=\sum_{k}\sigma^{(k)}_{z}$ and $J_{x}=\sum_{k}\sigma^{(k)}_{x}$ and the function
\begin{equation}
\textstyle\Xi(V,N_{g})=\left(1+\frac{V}{E_{C}}\right)(1-2N_{g}).
\end{equation}
Equation~(\ref{H2}) illustrates that multiple qubits behave in the same way if they are initialized in the same state. We define the dimensionless parameter $\eta=\frac{E^{2}_{C}}{E_{J}V}$ to measure the interqubit coupling strength.

\subsubsection{Ground state of nondissipative system}

\begin{figure}
\centering
\includegraphics[width=12.0cm]{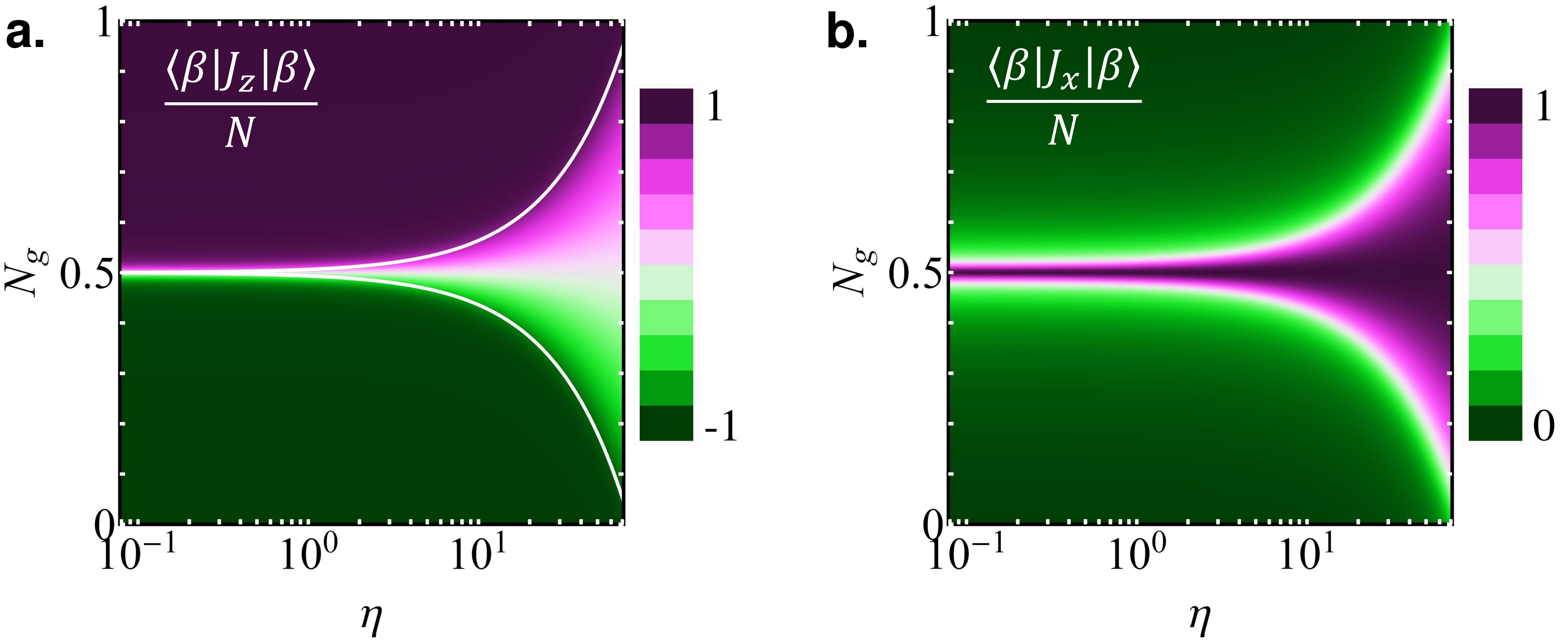}\\
\caption{(a) Ground-state expectation value $\frac{\langle\beta|J_{z}|\beta\rangle}{N}$ of nondissipative homogeneous circuit as a function of $\eta$ and $N_{g}$. Solid lines: $\Xi(V,N_{g})=\pm1$. The varying range of $\eta$ ensures the charging limit of $\frac{E_{C}}{E_{J}}\gg1$. (b) Dependence of expectation value $\frac{\langle\beta|J_{x}|\beta\rangle}{N}$ on $\eta$ and $N_{g}$.}\label{Fig2}
\end{figure}

We are interested in the ground state of the nondissipative system in the limit of $N\rightarrow\infty$. The Holstein-Primakoff transformation~\cite{PRA:Yu2014},
\begin{eqnarray}
\textstyle\frac{J_{z}}{2}&=&b^{\dag}b-\frac{N}{2},\\
\textstyle\frac{J_{x}}{2}&=&\sqrt{N-b^{\dag}b}b+b^{\dag}\sqrt{N-b^{\dag}b},
\end{eqnarray}
is employed to map the multiple-qubit system onto a bosonic mode. The annihilation $b$ and creation $b^{\dag}$ operators fulfill the bosonic commutation relation $[b,b^{\dag}]=1$. Then, we introduce the macroscopic displacements, i.e., $b\rightarrow b+\sqrt{N}\beta$ and $b^{\dag}\rightarrow b^{\dag}+\sqrt{N}\beta^{\ast}$, to this bosonic mode and obtain a displaced Hamiltonian 
\begin{equation}
{\cal{H}}=e^{-\sqrt{N}(\beta b^{\dag}-\beta^{\ast}b)}He^{\sqrt{N}(\beta b^{\dag}-\beta^{\ast}b)}.
\end{equation}
${\cal{H}}$ is further expanded as ordered power series in $b$ and $b^{\dag}$. In the second-order approximation, choosing $\beta$ such that the linear terms associated with $b$ and $b^{\dag}$ vanish leads to the ground state $|\beta\rangle$ and the corresponding energy
\begin{eqnarray}
\nonumber\textstyle{\frac{{\cal{E}}}{N}}&\equiv&\textstyle{\frac{\langle\beta|{\cal{H}}|\beta\rangle}{N}}\\
&=&\textstyle\frac{E^{2}_{C}}{4V}\left[2\beta^{2}-1+\Xi(V,N_{g})\right]^{2}-2E_{J}\sqrt{1-\beta^{2}}\beta.
\end{eqnarray}
Finally, we find that $\beta$ is real and determined by
\begin{equation}
\textstyle\frac{E^{2}_{C}}{V}\left[2\beta^{2}-1+\Xi(V,N_{g})\right]=E_{J}\frac{1-2\beta^{2}}{\beta\sqrt{1-\beta^{2}}},
\end{equation}
from which multiple solutions may be derived and the accepted one should minimize ${\cal{E}}$.

We focus on the observable operator $J_{z}$. Figure~\ref{Fig2}(a) shows the ground-state expectation value $\frac{\langle\beta|J_{z}|\beta\rangle}{N}=(2\beta^{2}-1)$ vs. $\eta$ and $N_{g}$. One can see that the large interqubit coupling ($\eta\gg1$) opens a wide intermediate regime, where $\frac{\langle\beta|J_{z}|\beta\rangle}{N}$ strongly relies on the gate-charge bias $N_{g}$. In contrast, for $\eta\ll1$ the system returns to an ensemble of independent qubits, and $\frac{\langle\beta|J_{z}|\beta\rangle}{N}\approx1$ ($-1$) when $N_{g}>\frac{1}{2}$ ($<\frac{1}{2}$). We should note that, unlike the Dicke model of a single cavity mode interacting with an ensemble of two-level atoms~\cite{PRL:Emary2003,PRE:Emary2003}, both first and second derivatives of ${\cal{E}}$ with respect to $\beta$ are continuous, indicating the absence of phase transitions in our system. In the limit of $E_{J}\sim0$, one obtains the simple solution, $\frac{\langle\beta|J_{z}|\beta\rangle}{N}=-\Xi(V,N_{g})$ and ${\cal{E}}=0$. Due to the condition of $-1\leq\frac{\langle\beta|J_{z}|\beta\rangle}{N}\leq1$, the ground-state diagram may be divided into three regimes: $\frac{\langle\beta|J_{z}|\beta\rangle}{N}=-1$ and $|\beta=0\rangle$ (all qubits are in $|0\rangle$), $\frac{\langle\beta|J_{z}|\beta\rangle}{N}=-\Xi(V,N_{g})$, and $\frac{\langle\beta|J_{z}|\beta\rangle}{N}=1$ and $|\beta=1\rangle$ (all qubits are in $|1\rangle$), and the boundaries are given by $\Xi(V,N_{g})=\pm1$ [see figure~\ref{Fig2}(a)]. We also displays the expectation value $\frac{\langle\beta|J_{x}|\beta\rangle}{N}$ in figure~\ref{Fig2}(b), where $\frac{\langle\beta|J_{x}|\beta\rangle}{N}$ always maximizes at the sweet spot $N_{g}=\frac{1}{2}$ due to the resonant driving.

\subsubsection{Dissipative system}

In the practical operation, the many-qubit system is unavoidably interfered by environmental fluctuations. It has been experimentally demonstrated that the dominant noise sources in a single charge qubit include the high-frequency Ohmic dissipation from the DC voltage source and the low-frequency $1/f$ noise induced by background charge fluctuations~\cite{PRL:Astafiev2004,PRL:Nakamura2012,Pashkin2009}. The former mainly determines the relaxation time $T_{1}$ of the charge qubit, i.e., the characteristic time scale of the damped qubit Rabi oscillation, while the latter primarily affects the dephasing time $T_{2}$, i.e., the characteristic time scale of the damped qubit Ramsey/spin echo oscillation. According to~\cite{Pashkin2009,Makhlin2003}, the whole noise may be mapped onto the gate-charge bias $N_{g}$, i. e., the voltage source, in the mathematical treatment. This is still valid in this multi-qubit scheme. Since all islands are electrostatically biased by the same voltage source, different qubits are subjected to the same Ohmic noise whose spectrum is proportional to the noise frequency $f$. In addition, all islands are strongly coupled in the system. When environmental fluctuations interrupt the dynamic of one Cooper-pair box, the local voltage of the corresponding gate capacitor is disturbed. This voltage fluctuation may be mapped onto the voltage source and it further influences the dynamics of other boxes in the same way, leading to the energy relaxation of other charge qubits. Thus, one may map all local voltage fluctuations occurring in different Cooper-pair boxes on to the single voltage source. Moreover, in this work we mainly focus on the qubit Rabi oscillation, whose damping time is $T_{1}$, and the low-frequency $1/f$ noise in the circuit hardly influences $T_{1}$. Therefore, for these reasons it is valid to model environmental fluctuations by the fluctuation of gate-charge bias (voltage source), i.e., all charge qubits suffer the same noise. This differs from the many-body system composed of excited-state atoms that spontaneously decay independently.

\begin{figure}
\centering
\includegraphics[width=13.0cm]{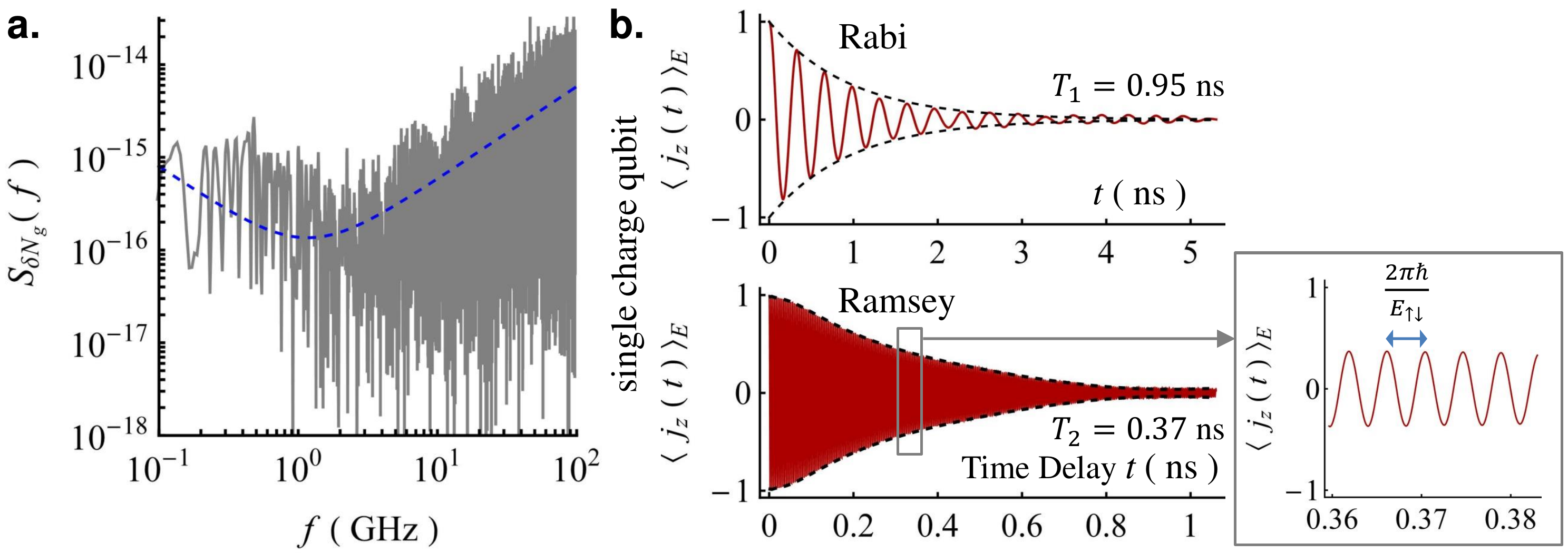}\\
\caption{(a) Power spectral density $S_{\delta N_{g}}(f)$ of gate-charge fluctuations $\delta N_{g}(t)$. Solid (dash) line corresponds to the numerical (analytical) result. (b) Single-charge-qubit Rabi and Ramsey oscillations with the system parameters of $C_{g}$, $C_{j}$, and $E_{J}$ same to figure~(\ref{Fig1}) and $E_{C}=78E_{J}$ for $C_{c}=0$. For the Rabi oscillation, the qubit is initialized in $|1\rangle$ with $N_{g0}=\frac{1}{2}$. For performing Ramsey fringes, the qubit is initially prepared in $|1\rangle$ and $N_{g0}$ is set at $\frac{1}{2}$ during two $\frac{\pi}{2}$-pulses while $N_{g0}=0$ in the free evolution period. Solid curves: numerical results. Dash lines: decay-envelope fittings. In the Ramsey oscillation, the detailed behavior around the characteristic time sale $T_{2}$, which is surrounded by the rectangular frame, is zoomed in, and the oscillation frequency is given by $E_{\uparrow\downarrow}=\sqrt{E^{2}_{C}(1-2N_{g})^{2}+E^{2}_{J}}$. For all curves of ensemble average $\langle j_{z}(t)\rangle_{E}$, the size of trajectory ensemble is $10^{5}$.}\label{Fig3}
\end{figure}

We rewrite the gate-charge bias as
\begin{equation}
N_{g}(t)=N_{g0}+\delta N_{g}(t),
\end{equation}
i.e., a constant value $N_{g0}$ plus a fluctuating term $\delta N_{g}(t)$. The noise spectral density
\begin{equation}
S_{\delta N_{g}}(f)=\textstyle\int^{\infty}_{-\infty}\int^{\infty}_{-\infty}\delta N_{g}(t+\tau)\delta N_{g}(t)e^{-i2\pi f\tau}dtd\tau,
\end{equation}
is given by
\begin{equation}
S_{\delta N_{g}}(f)=\textstyle\frac{\pi R\hbar C^{2}_{g}}{e^{2}}f+\frac{\alpha}{2\pi f},
\end{equation}
where the typical impedance of the voltage-source circuit is $R=50$ $\Omega$ and $\alpha=5.0\times10^{-7}$~\cite{APL:Zimmerli1992,JAP:Verbrugh1995,IEEE:Wolf1997}. Accordingly to $S_{\delta N_{g}}(f)$, one may numerically generate $\delta N_{g}(t)$ [see figure~\ref{Fig3}(a)].

We choose $\{|n_{1}\rangle\otimes\cdots\otimes|n_{N}\rangle;n_{k=1,\ldots,N}=0,1\}$ to span the Hilbert space. Using the Schr\"{o}dinger equation,
\begin{equation}
\textstyle i\hbar\frac{d}{dt}\psi(t)=H\psi(t),
\end{equation}
one can simulate the system's state $\psi(t)$ for a given initial state $\psi(0)$, resulting in the trajectory
\begin{equation}
j_{z}(t)=\langle\psi(t)|J_{z}|\psi(t)\rangle/N.
\end{equation}
Repeating the simulation with the same initial condition leads to the ensemble mean observation $\langle j_{z}(t)\rangle_{E}$. Here we use $\langle\ldots\rangle_{E}$ to denote the ensemble average. As an example, figure~\ref{Fig3}(b) depicts the Rabi and Ramsey oscillations of a single charge qubit, from which the decoherence times $T_{1,2}$ may be extracted.

\begin{figure}
\centering
\includegraphics[width=11.0cm]{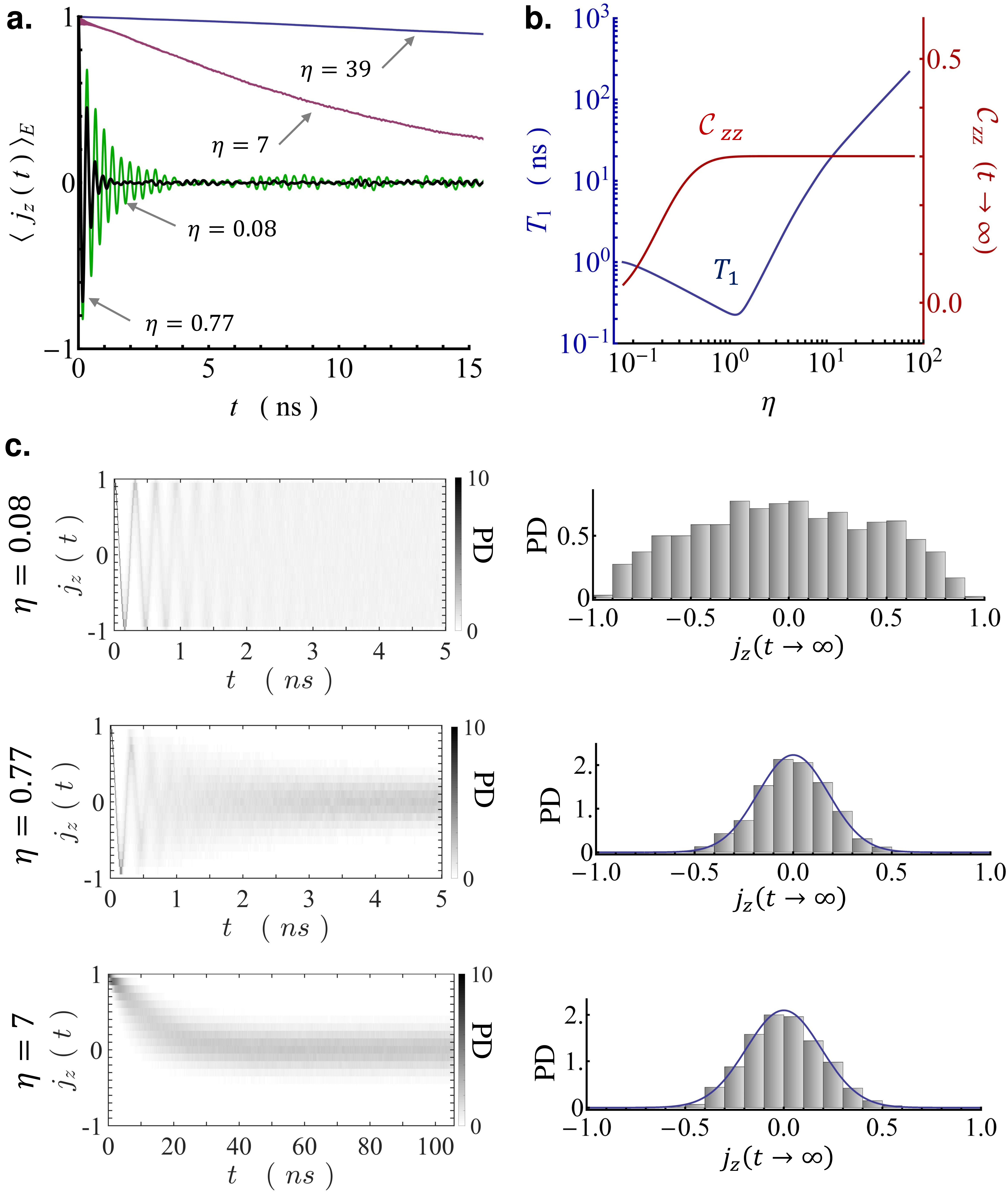}\\
\caption{(a) Rabi oscillation $\langle j_{z}(t)\rangle_{E}$ of the dissipative homogeneous system where all charge qubits are initialized in $|n_{k=1,\ldots,N}=1\rangle$. The relaxation time $T_{1}$ is extracted from the decay envelope. (b) $T_{1}$ and steady-state ${\cal{C}}_{zz}(t\rightarrow\infty)$ vs. $\eta$. (c) Time-dependent probability density (PD) distribution of the trajectory ensemble of $j_{z}(t)$ and histogram of steady-state ensemble of $j_{z}(t\rightarrow\infty)$ for several different $\eta$, where the solid curves correspond to the line fittings. For $\eta=0.08$, $0.77$, and $7$, the corresponding standard deviations $\Delta j_{z}(t\rightarrow\infty)$ are derived as $0.45$, $0.18$, and $0.18$, respectively. For all curves, the system parameters are same to figure~\ref{Fig1} and the ensemble size is chosen to be $10^{3}$.}\label{Fig4}
\end{figure}

In the following we only focus on the collective Rabi oscillation of multiple-qubit circuit, where the system is initialized in $|1\rangle_{1}\otimes\ldots\otimes|1\rangle_{N}$ and we set $N_{g0}=\frac{1}{2}$. Figure~\ref{Fig4}(a) shows the damped $\langle j_{z}(t)\rangle_{E}$ for several different $\eta$. It is seen that for $\eta\ll1$, i.e., the very weak qubit-qubit interaction, $\langle j_{z}(t)\rangle_{E}$ is similar to that of single qubit, meaning that multiple qubits act almost independently. When $\eta$ is increased, i.e., the interqubit coupling becomes strong, the relaxation time $T_{1}$ of $\langle j_{z}(t)\rangle_{E}$ is dramatically reduced because the qubits get correlated and their dynamics are affected with each other. When one qubit relaxes, the gate-charge bias is disturbed and this influences the dynamics of other qubits, leading to an enhanced collective relaxation. As $\eta$ is further increased, the interqubit interactions become very strong.

Surprisingly, $\langle j_{z}(t)\rangle_{E}$ relaxes much more slowly than that of single qubit and the oscillation behavior of $\langle j_{z}(t)\rangle_{E}$ also disappears. It is understandable from the aspect of interaction-induced detunings~\cite{JModOpt:Yu2016}. Diagonalizing the Hamiltonian~(\ref{H2}) leads to the eigenstates of the system. For the noninteracting system ($C_{c}=0$), the eigenstates with the same total number of the qubits in $|1\rangle$ are degenerate. The interqubit coupling removes this degeneracy and gives rise to the energy-level shifts of eigenstates. The much strong interactions among qubits enhance the qubit-state shifts to the values comparable or even larger than the Josephson energy $E_{J}$. Thus, the system operating point moves far away from the sweet spot $N_{g}=\frac{1}{2}$, resulting in a large qubit detuning and the suppression of oscillation behavior of $\langle j_{z}(t)\rangle_{E}$. In addition, according to the relaxation-time formula derived from the Fermi's golden rule~\cite{RMP:Makhlin2001}, $T_{1}$ is extended when the system moves away from the optimal working point.

The dependence of $T_{1}$ on $\eta$ is displayed in figure~\ref{Fig4}(b). It is shown that $T_{1}$ starts to rise after $\eta\approx1$ where, as illustrated by equation~(\ref{H2}), the energy-level shift (detuning) $\frac{E^{2}_{C}}{V}$ is equal to the driving strength $E_{J}$. We should point out that although $T_{1}$ is enhanced in the large interqubit-coupling regime, the qubit dephasing time $T_{2}$ is strongly suppressed. At the sweet spot $N_{g}=\frac{1}{2}$, the qubit is minimally sensitive to the $1/f$ fluctuation in the quantum circuit. When the qubit working point departs from $N_{g}=\frac{1}{2}$, the $1/f$ noise significantly reduces $T_{2}$~\cite{PRL:Astafiev2004,PRL:Yoshihara2006,PRA:Koch2007}. Nevertheless, the influence of the $1/f$ noise may be weakened via stabilizing the superconducting qubit to a high-$Q$ resonator~\cite{Nature:Vijay2012} and the feedback control method~\cite{NPJ:Yu2018,PRA:Yu2018}, which potentially extends the dephasing time $T_{2}$.

After a long enough time $(t\gg T_{1})$, the multiple-qubit system loses the memory of its initial condition and the ensemble expectation value $\langle j_{z}(t)\rangle_{E}$ approaches a steady-state value~\cite{JOSAB:Molmer1993,RMP:Plenio1998}. We use the symbol $t\rightarrow\infty$ to denote a large time scale after which the distribution of $j_{z}(t)$ becomes steady, i.e., the ensemble-averaged observables reach steady-state values. As shown in figure~\ref{Fig4}(a), $\langle j_{z}(t\rightarrow\infty)\rangle_{E}$ is equal to zero and independent of the coupling parameter $\eta$. Nevertheless, the steady-state histogram of the ensemble of $j_{z}(t\rightarrow\infty)$ relies on $\eta$. As depicted in figure~\ref{Fig4}(c), the trajectory ensemble of $j_{z}(t\rightarrow\infty)$ is distributed over the entire range from -1 to 1 in the weak-coupling limit ($\eta\ll1$). In contrast, as the qubit-qubit interaction is increased, although the qubits are still equally populated on $|0\rangle$ and $|1\rangle$, the spread of the statistical distribution of $j_{z}(t\rightarrow\infty)$ is narrowed, indicating the reduction of measurement uncertainty and the enhanced interqubit correlation. Generally, the expectation value of a product of two operators ($O_{1}$ and $O_{2}$) is different from the product of the expectation values of individual operators, i.e., $\langle\psi|O_{1}O_{2}|\psi\rangle\neq\langle\psi|O_{1}|\psi\rangle\langle\psi|O_{2}|\psi\rangle$, which is attributed to the correlation between two quantities. Accordingly, since the qubits are coupled via the $zz$-interaction [see equation~(\ref{H1b})], we define
\begin{equation}
{\cal{C}}_{zz}(t)=\textstyle{\left\langle\frac{1}{N(N-1)}\sum_{k_{1}\neq k_{2}}\left[\langle\psi(t)|\sigma^{(k_{1})}_{z}\sigma^{(k_{2})}_{z}|\psi(t)\rangle-\langle\psi(t)|\sigma^{(k_{1})}_{z}|\psi(t)\rangle\langle\psi(t)|\sigma^{(k_{2})}_{z}|\psi(t)\rangle\right]\right\rangle_{E}},
\end{equation}
to measure the total interqubit correlation in the system~\cite{NewJPhys:Schachenmayer2015}. For independent qubits one has ${\cal{C}}_{zz}(t\rightarrow\infty)=0$. As exhibited in figure~\ref{Fig4}(b), ${\cal{C}}_{zz}(t\rightarrow\infty)$ approximates zero in the limit of $\eta\ll1$, indicating that the qubits behave independently. As $\eta$ is increased, ${\cal{C}}_{zz}(t\rightarrow\infty)$ goes up strongly and is saturated eventually.

We should note that ${\cal{C}}_{zz}(t)$ differs from the variance $\Delta j^{2}_{z}(t)$ of the ensemble of trajectories $j_{z}(t)$, the definition of which is $\Delta j^{2}_{z}(t)=\left\langle j^{2}_{z}(t)\right\rangle_{E}-\left\langle j_{z}(t)\right\rangle_{E}^{2}$. $\Delta j_{z}(t)$ weights the ensemble spread of $j_{z}(t)$. Figure~\ref{Fig4}(c) shows the time-dependent ensemble distribution of $j_{z}(t)$ and steady-state histogram of $j_{z}(t\rightarrow\infty)$ for several different $\eta$. The distribution of $j_{z}(t)$ diffuses rapidly when $\eta$ is increased from zero, corresponding to the strong reduction of $T_{1}$. However, the width of steady-state distribution of $j_{z}(t\rightarrow\infty)$ becomes narrow, i.e., $\Delta j_{z}(t\rightarrow\infty)$ is suppressed, meaning that the interqubit correlation is enhanced. For $\eta\gg1$, the ensemble of $j_{z}(t)$ spreads slowly [see $j_{z}(t)$ with $\eta=0.77$ and $\eta=7$ in figure~\ref{Fig4}(c)] and the histogram of $j_{z}(t\rightarrow\infty)$ barely changes compared with that of $\eta\approx1$, denoting the extension of $T_{1}$ and the saturation of ${\cal{C}}_{zz}(t\rightarrow\infty)$.

Actually, the spreading rate of the histogram of $j_{z}(t)$ corresponds to the relaxation rate of the multi-qubit system. As shown in figure~\ref{Fig4}(c), the system is initially prepared in the ground state $|1\rangle_{1}\otimes\ldots\otimes|1\rangle_{N}$. The histogram width of the trajectory ensemble $j_{z}(t)$ at $t=0$ is zero in principle. As the time $t$ is increased, the histogram of $j_{z}(t)$ becomes broader due to the fluctuation in the quantum circuit. For the system with small (large) $T_{1}$, the histogram of $j_{z}(t)$ reaches the steady-state distribution fast (slowly).

\subsection{Inhomogeneous Circuit}

So far, we have only focused on the homogeneous system with identical qubits. However, in the practical circumstance it is extremely difficult to fabricate an array of identical Josephson junctions. Thus, we have to return to the inhomogeneous system described by equations~(\ref{H1a}) and (\ref{H1b}). To take into account this nonideal-fabrication-induced inhomogeneity, we assume a normal distribution with a standard deviation $\lambda$ for different junction areas whose mean value is normalized to be unity~\cite{PRL:Kakuyanagi2016}. We use $C_{j}$ and $E_{J}$ to respectively denote the mean values of self-capacitances $C_{j,k=1,\ldots,N}$ and Josephson energies $E_{J,k=1,\ldots,N}$, i.e., $C_{j}=\frac{1}{N}\sum_{k}C_{j,k}$ and $E_{J}=\frac{1}{N}\sum_{k}E_{J,k}$. The corresponding standard deviations are given by $\lambda C_{j}$ and $\lambda E_{J}$ since both $C_{j,k}$ and $E_{J,k}$ are proportional to the area of the $k$-th junction. For the charging energies $E_{C,k=1,\ldots,N}$, the mean value is $E_{C}=\frac{(2e)^{2}}{2C_{\Sigma}}$ with $C_{\Sigma}=C_{g}+C_{j}+C_{c}$ and the standard deviation is derived as $\lambda\frac{C_{j}}{C_{\Sigma}}E_{C}$ in the limit of $C_{\Sigma}\gg C_{j}$. In addition, the ensemble average $\langle\ldots\rangle_{E}$ needs to involve this extra inhomogeneity, for which we assume that a group of $N$-qubit systems are prepared in the same initial condition and $E_{C,k=1,\ldots,N}$ and $E_{J,k=1,\ldots,N}$ of each system fulfill the corresponding normal distributions.

\begin{figure}
\centering
\includegraphics[width=12.0cm]{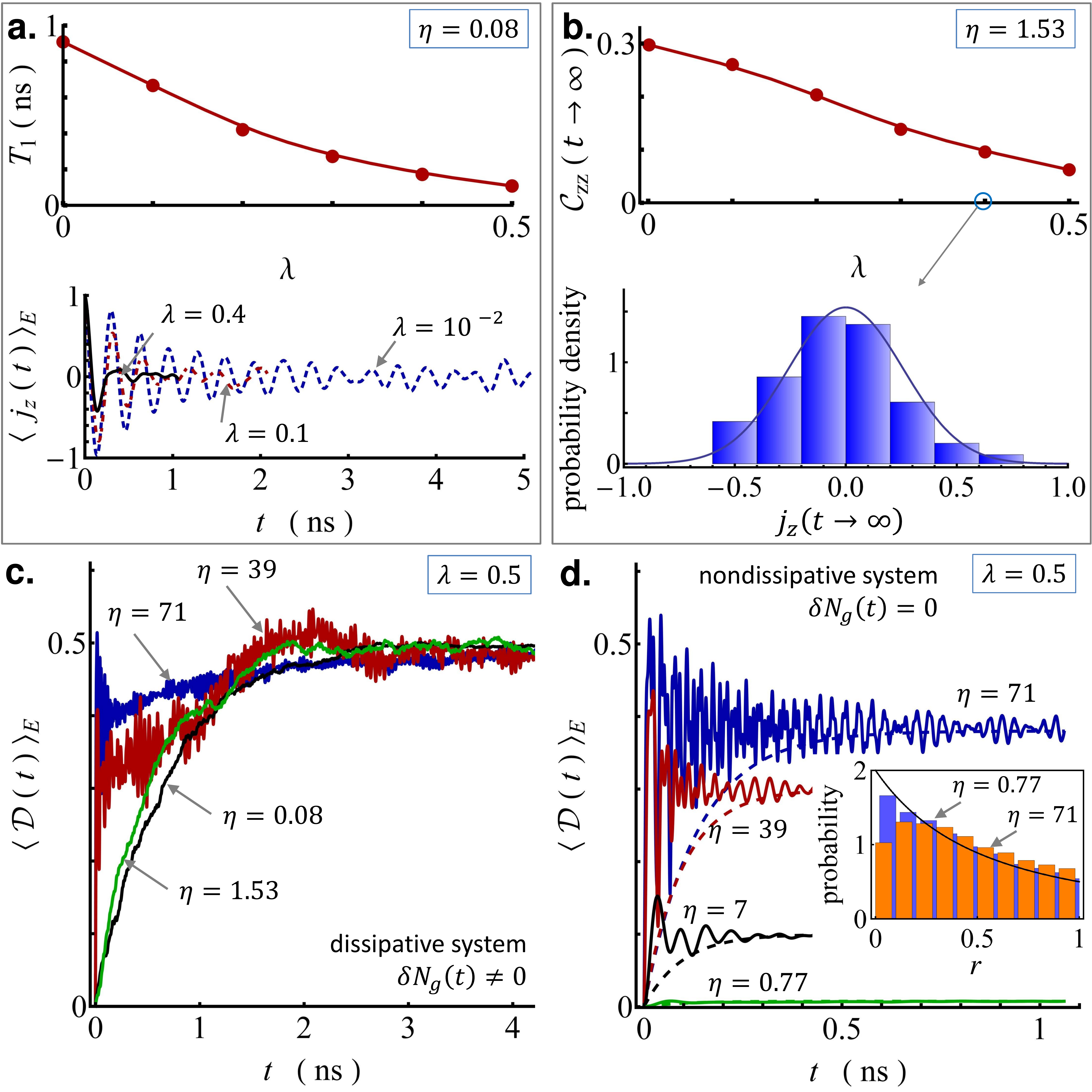}\\
\caption{Inhomogeneous multiple-qubit system and its application in many-body localization. (a) Upper: Relaxation time $T_{1}$ vs. standard deviation $\lambda$ with $\eta=0.08$. Lower: Rabi oscillation $\langle j_{z}(t)\rangle_{E}$ for several different $\lambda$. (b) Upper: ${\cal{C}}_{zz}(t\rightarrow\infty)$ as a function of $\lambda$ with $\eta=1.53$. Lower: Histogram of $j_{z}(t\rightarrow\infty)$ with $\lambda=0.4$ and $\Delta j_{z}(t\rightarrow\infty)=0.26$. (c) $\langle{\cal{D}}(t)\rangle_{E}$ of the disordered dissipative system with $N_{g0}=\frac{1}{2}$ and $\lambda=0.5$. (d) $\langle{\cal{D}}(t)\rangle_{E}$ of the disordered nondissipative system with $N_{g}=\frac{1}{2}$ and $\lambda=0.5$. The dashed lines are the results from the curve fitting based on $\langle{\cal{D}}(t\rightarrow\infty)\rangle_{E}(1-e^{-\gamma t})$. The localization rate approximates $\gamma\approx2\pi\times1.6$ GHz for different $\eta$. The values of $\langle{\cal{D}}(t\rightarrow\infty)\rangle_{E}$ are about 0.01, 0.1, 0.3, and 0.38 for $\eta=0.77$, 7, 39, and 71, respectively. Inset: Histogram of the ratio $r$ of adjacent energy-level gaps, where the solid curve corresponds to the Poisson distribution. The ensemble size is chosen to be $10^{3}$ for all curves. All system parameters are same to figure~\ref{Fig1}.}\label{Fig5}
\end{figure}

As shown in figure~\ref{Fig5}(a), the inhomogeneity caused by nonidentical Josephson junctions strongly accelerates the relaxation of the collective Rabi oscillation even in the weak-coupling limit ($\eta\ll1$). This corresponds to the inhomogeneous broadening, where the unsynchronized Rabi dynamics of individual qubits destructively interfere with each other. Moreover, the extra inhomogeneity reduces the interqubit correlation and broadens the trajectory distribution [figure~\ref{Fig5}(b)]. Therefore, maximally minimizing the inhomogeneity arising from the defective fabrication technology becomes indispensable for the application of multiple-qubit circuits in QIP.

Nevertheless, this inhomogeneous system may be still potentially applied to study the quantum many-body localization. We choose two spin states of the $k$-th qubit as $|\uparrow\rangle_{k}=\frac{1}{\sqrt{2}}(|0\rangle_{k}+|1\rangle_{k})$ and $|\downarrow\rangle_{k}=\frac{1}{\sqrt{2}}(|0\rangle_{k}-|1\rangle_{k})$ and define the new $x$- and $y$-components of Pauli matrix as $\tilde{\sigma}_{x,k}\equiv\sigma_{z,k}$ and $\tilde{\sigma}_{z,k}\equiv\sigma_{x,k}$. Then, the inhomogeneous Hamiltonian $H$ can be mapped onto the Ising model~\cite{NatPhys:Smith2016}
\begin{equation}
H_{Ising}=\textstyle\sum_{k<k'}J_{k,k'}\tilde{\sigma}_{x,k}\tilde{\sigma}_{x,k'}-\frac{B}{2}\sum_{k}\tilde{\sigma}_{z,k}-\sum_{k}\frac{D_{k}}{2}\tilde{\sigma}_{z,k}+\textstyle\sum_{k}\Delta_{k}\tilde{\sigma}_{x,k},
\end{equation}
where $J_{k,k'}=\frac{E_{C,k}E_{C,k'}}{4NV}$ corresponds to the long-range spin-spin interaction, $B=E_{J}$ plays the role of the external field, and the inhomogeneity $D_{k}=E_{J,k}-E_{J}$ acts as the site-dependent disordered potential. Unlike the superconducting circuit in~\cite{arXiv:Xu2017}, the nearest-neighbor $J_{k,k\pm1}$ do not dominate. The last term in $H_{Ising}$ with
\begin{equation}
\Delta_{k}=\textstyle\frac{E_{C,k}}{2}(1-2N_{g})\left(1+\sum_{k'}\frac{E_{C,k'}}{NV}\right),
\end{equation}
vanishes at the sweet point $N_{g}=\frac{1}{2}$. However, the nonzero $\delta N_{g}(t)$ deviates $H_{Ising}$ from the standard disordered Ising model. In addition, the disorder term $D_{k}$ is sampled from a normal distribution, rather than a uniform random variable~\cite{arXiv:Xu2017}.

According to~\cite{NatPhys:Smith2016}, in the weak-coupling limit ($\eta\ll1$), we have $\lambda E_{J}\gg\textrm{max}(J_{k,k'})$ and the system may emerge the multiple-spin localization, which is quantified by the normalized Hamming distance~\cite{PRB: Hauke2015}
\begin{equation}
{\cal{D}}(t)=\textstyle\frac{1}{2}-\frac{1}{2N}\sum_{k}\langle\psi(0)|e^{i\frac{H_{Ising}}{\hbar}t}\tilde{\sigma}_{z,k}e^{-i\frac{H_{Ising}}{\hbar}t}\tilde{\sigma}_{z,k}|\psi(0)\rangle,
\end{equation}
with the initial state $\psi(0)$ being the N\'{e}el state. ${\cal{D}}(t\rightarrow\infty)$ arrives at $\frac{1}{2}$ for a thermalizing state and remains 0 at a fully localized state. For the disordered dissipative system with ($\lambda\neq0$, $\delta N_{g} (t)\neq0$), the Hamming distance ${\cal{D}}(t\rightarrow\infty)$ always approaches $\frac{1}{2}$, reaching the thermal equilibrium [see figure~\ref{Fig5}(c)]. This is attributed to the effect of environmental fluctuations $\delta N_{g} (t)$. Thus, besides QIP, suppressing the environmental noise is also the key issue for the application of superconducting circuits in the many-body simulation. Figure~\ref{Fig5}(d), where we have artificially set $\delta N_{g}(t)=0$, exhibits that the disordered nondissipative system with ($\lambda\neq0$, $\delta N_{g} (t)=0$) stays nearly fully localized product states in the weak-coupling limit ($\eta\ll1$) while ${\cal{D}}(t\rightarrow\infty)$ rises gradually up to $\frac{1}{2}$ as the qubit-qubit interactions become strong. Using the diagonalization method, we have also checked the statistics of the ratio parameter $r$ of adjacent energy-level gaps for the disordered nondissipative system~\cite{PRB:Oganesyan2007,PRB:Pal2010}. It is seen that $r$ follows the Poisson distribution when $\eta\ll1$ [the inset of figure~\ref{Fig5}(d)], manifesting the many-body localization, and apparently violates the Poisson distribution for a larger $\eta$, suppressing the localization effect.

As experimentally demonstrated in~\cite{arXiv:Xu2017}, the many-body localized state may be still attainable within a short time scale when the thermalization rate of the system with ($\lambda=0$, $\delta N_{g} (t)\neq0$) coupling to the environment is much slower than the localization rate of the system with ($\lambda\neq0$, $\delta N_{g} (t)=0$) (i.e., the localization rate measures how fast a disordered nondissipative system reaches the localized state from an initially-prepared state). The localization rate $\gamma$ of the system with ($\lambda\neq0$, $\delta N_{g} (t)=0$) may be roughly estimated by assuming that the envelop of ${\cal{D}}(t)$ follows the exponential law, i.e., $\langle{\cal{D}}(t\rightarrow\infty)\rangle_{E}(1-e^{-\gamma t})$ [see the dashed lines in figure~\ref{Fig5}(d)]. The time scale that measures the thermalization rate of the system with ($\lambda=0$, $\delta N_{g} (t)\neq0$) coupling to the environment is given by $T_{1}(\lambda=0)$. The many-body localized state is potentially observed in the disordered dissipative system when $\gamma\gg T^{-1}_{1}(\lambda=0)$ which may be fulfilled for a large $\eta$.

In comparison between figure~\ref{Fig5}(c) and~\ref{Fig5}(d), we find that for $\eta\gg1$ the Hamming distance ${\cal{D}}(t)$ of the disordered dissipative system rapidly reaches a metastable value smaller than $\frac{1}{2}$ (i.e., the system arrives at the localized state) and then slowly grows up to $\frac{1}{2}$ (i.e., the system approaches the thermalizing state). Thus, the many-body localization phenomenon can be observed within a short time scale for $\eta\gg1$. Increasing the disorder strength $D_{k}$, i.e., increasing $\lambda$, may enhance the localization rate of the inhomogeneous system, leading to an easy access to the many-body localized phase. However, a large $\lambda$ strongly reduces $\frac{E_{C,k}}{E_{J,k}}$, i.e., some qubits may not operate in the charging limit ($\frac{E_{C,k}}{E_{J,k}}\gg1$).

\section{Conclusion}

We have studied a multiple-charge-qubit system, where the Cooper-pair boxes are all capacitively linked. The qubit dynamics are interfered with environmental fluctuations which are mapped onto the gate voltage source that biases all qubits. The collective Rabi oscillation has been numerically simulated for both homogeneous and inhomogeneous systems. We find the interqubit coupling strongly varies the energy-relaxation rate of the quantum circuit consisting of identical Josephson junctions. For the weak coupling system, the qubit relaxation time $T_{1}$ is significantly reduced since the dynamics of one qubit is unavoidably influenced by the fluctuations of other qubits. In contrast, $T_{1}$ of the homogeneous system in the strong coupling regime can be enhanced to a value much larger than that of a free qubit. This result is caused by the strong energy-level shifts of qubit states (i.e., the large interaction-induced detunings) and consistent with the expectation of Fermi's golden rule. In QIP, transferring quantum information between two qubits in a homogeneous multi-qubit network is an essential process. The resulting fidelity is limited by the system's decoherence time. Thus, the enhancement of $T_{1}$ is crucially relevant to QIP.

The nonideal-fabrication-induced inhomogeneity always expedites the multi-qubit system's collective decay. In addition, we mapped the inhomogeneous system onto the disordered Ising model to probe the many-body localization effect. This enables us to investigate the role of the environmental noise in the quantum simulation. For the inhomogeneous system with the localization rate faster than the thermalization rate, the many-body-localization regime is still accessible.

\section{Acknowledgements}
This research has been supported by the National Research Foundation Singapore \& by the Ministry of Education Singapore Academic Research Fund Tier 2 (Grant No. MOE2015-T2-1-101).

\end{document}